\documentclass[12pt,a4paper]{article}
\usepackage{amsfonts,amssymb,amsmath,amsthm,graphicx,psfrag,bbm}
\usepackage[english]{babel}
\newtheorem{thm}{Theorem}

\newtheorem{lem}[thm]{Lemma}

\newcommand{\Z}{\mathbb{Z}}

\newcommand{\R}{\mathbb{R}}
\newcommand{\st}{\textrm{ s.t. }}
\newcommand{\Pb}{\mathbb{P}}
\newcommand{\e}{\varepsilon}
\newcommand{\eone}{\textrm{e}}
\newcommand{\setm}{\setminus}
\newcommand{\Or}{\mathcal{O}}

\newcommand{\Dtt}[2]{\frac{\ddritta^2 #1}{\ddritta \! #2^2}}
\DeclareMathOperator*{\Det}{Det}
\DeclareMathOperator*{\ddritta}{d}

\author{Patrik L. Ferrari and Herbert Spohn\\
{\normalsize Zentrum Mathematik, TU M\"unchen}\\{\normalsize D-85747 Garching, Germany}\\
{\normalsize emails: ferrari@ma.tum.de, spohn@ma.tum.de}}
\date{}

\begin{document}

\title{Last Branching in Directed Last Passage Percolation}
\maketitle

\begin{abstract}
The $1+1$ dimensional directed polymers in a Poissonean random environment is
studied. For two polymers of maximal length with the same origin and
distinct end points we establish that the point of last branching is governed by
the exponent for the transversal fluctuations of a single polymer. We also
investigate the density of branches.
\end{abstract}

{\sc Running Title: }Last Branching

\section{Introduction and main result}

First passage percolation was invented as a simple model for the spreading of a
fluid in a porous medium. One imagines that the fluid is injected at the origin.
Upon spreading the time it takes to wet across a given bond is postulated to be
random. In the directed version the wetting is allowed along a preferred
direction only. The task is then to study the random shape of the wetted region
at some large time $t$. The existence of a deterministic shape as $t\to\infty$
follows from the subadditive ergodic theorem~\cite{Kesten}. The shape
fluctuations are more difficult to analyse and only some bounds are
available~\cite{Piza}.

A spectacular progress has been achieved recently by Baik, Deift, and
Johansson~\cite{BDJ}, who prove that for directed first passage percolation in
two dimensions the wetting time measured along a fixed ray from the origin has fluctuations of
order $t^{1/3}$. The amplitude has a non-Gaussian distribution. In fact it is
Tracy-Widom distributed~\cite{TW}, a distribution known previously from the theory of Gaussian
random matrices. Of course, such a detailed result is available only for a very
specific model. In this model the
wetting time is negative, which can be converted into a positive one at the expense
of studying \emph{last} rather than first passage percolation, hence our title.
One thereby loses the physical interpretation of the spreading of a fluid. But
directed first and last passage percolation models are expected to be in the
same universality class under the condition that along the ray under
consideration the macroscopic shape has a non-zero
curvature~\cite{KMH,PS}.

Such detailed results are available only for a few last passage percolation
models, among them the Poissonean model studied in~\cite{BDJ}.
It was first introduced by Hammersley~\cite{Ha}, cf. also
the survey by Aldous and Diaconis~\cite{AD}. We start from a Poisson point process on
$\R_+^2$ with intensity one. Let $(x,y)\prec (x',y')$ if $x<x'$ and $y<y'$. 
For a given configuration $\omega$ of the Poisson process and two points
$S\prec E \in \R_+^2$ a \emph{directed polymer} starting at $S$ and ending at
$E$ is a piecewise linear path $\pi$ obtained by connecting $S$ and $E$ through
a subset $\{q_1,\ldots,q_N\}$ of points in $\omega$ such that $S\prec q_1 \prec \cdots
\prec q_N \prec E$. The length, $l(\pi)$, of the directed polymer
$\pi$ is the number of Poisson points visited by $\pi$. We denote by
$\Pi(S,E,\omega)$ the set of all directed polymers from $S$ to $E$ for given
$\omega$ and we are interested in directed polymers which have maximal length. In
general there will be several of these and we denote by $\Pi_{\max}(S,E,\omega)$
the set of maximizers, i.e. of directed polymers in $\Pi(S,E,\omega)$ with \emph{maximal length}
$$L(S,E)(\omega)= \max_{\pi \in \Pi(S,E,\omega)} l(\pi).$$

\begin{figure}[t]\label{simulation}
\begin{center}
\includegraphics[angle=0,width=16cm]{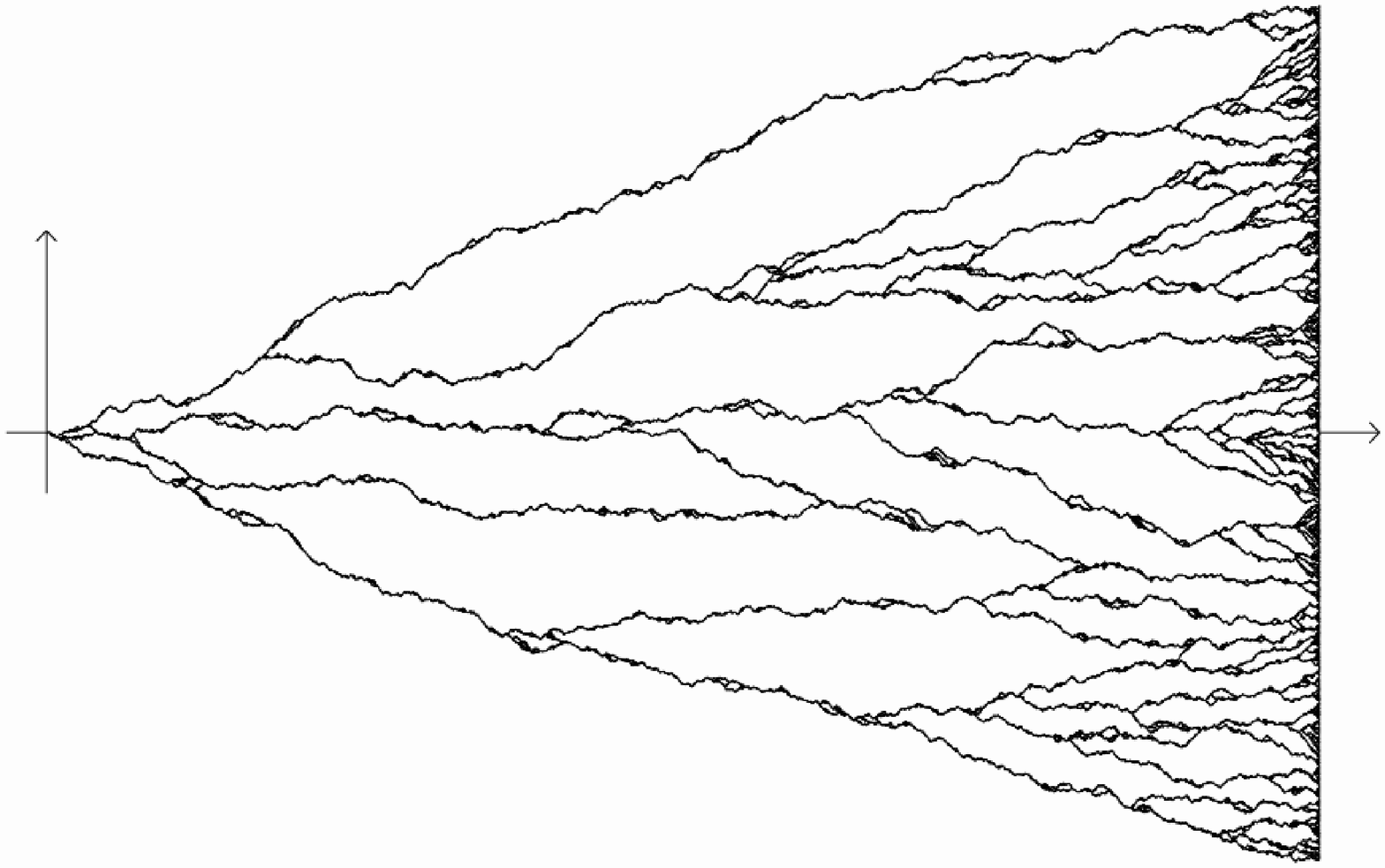}
\caption{\it Set of all maximizers from the origin to the line $U_t$. The
sample uses \mbox{$\sim 8\cdot 10^6$} Poisson points, which in our units correspond to
$t=2000$. Only the section \mbox{$[0,1]\times[-1/6,1/6]$} is shown.}
\end{center}
\end{figure}

For the specific choice $S=(0,0)$, $E=(t,t)$ let us set $L(S,E)=L(t)$.
The distribution function for $L(t)$ can be written in determinantal form as
\begin{equation}\label{eq1v2}
\Pb(L(t)<a)=\Det(\mathbbm{1}- P_a B_t)
\end{equation}
Here $P_a$ and $B_t$ are projection operators in $\ell_2(\Z)$. $P_a$ projects
onto $[a,\infty)$ and $B_t$ is the spectral projection corresponding to the
interval $(-\infty,0]$ of the operator $H_t$ defined through
\begin{equation}
H_t\, \psi(n)=-\psi(n+1)-\psi(n-1)+\frac{n}{t}\psi(n),
\end{equation}
i.e. $B_t$ is the discrete Bessel kernel. (\ref{eq1v2}) should be compared with
the determinantal formula for the largest eigenvalue, $E_{\max}$, of a $N\times
N$ Gaussian, $\beta=2$ random matrix, which has the distribution function
\begin{equation}
\Pb(E_{\max}\leq a)=\Det(\mathbbm{1}-P_a K_N).
\end{equation}
Here $P_a$ and $K_N$ are projections in $L^2(\R)$.
$P_a$ projects onto the
semiinfinite interval $[a,\infty)$ and $K_N$ is the spectral projection onto
the interval $[0,N]$ of the operator $-\frac{1}{2}\Dtt{}{x}+\frac{1}{2}x^2$. In
the limit of large $t$, under suitable rescaling \cite{PS,TW}, both determinantal formulae
converge to
\begin{equation}\label{eq4v2}
\Det(\mathbbm{1}-P_a K)
\end{equation}
where $K$ is the Airy kernel, i.e. the spectral projection corresponding to
$(-\infty,0]$ of the Airy operator $-\Dtt{}{x}+x$. (\ref{eq4v2}) is the
distribution function for a standard Tracy-Widom random variable
$\zeta_2$~\cite{TW}. The famous result in~\cite{BDJ} states that
\begin{equation}\label{eq5v2}
L(t)\cong 2t+t^{1/3}\zeta_2
\end{equation} in the limit of large $t$.
In brackets, we remark that the proof in~\cite{BDJ} proceeds via Toeplitz and
not, as indicated here, via Fredholm determinants.

While (\ref{eq5v2}) gives very precise information about the typical length of
directed polymer, it leaves untouched the issue of typical spatial excursions of a maximizing directed
polymer. As shown in~\cite{Jo}, they are in fact of size $t^{2/3}$ away from
the diagonal. No information on the distribution is available. The transverse
exponent $2/3$ appears also in a somewhat different quantity~\cite{PS}. Set
$S=(0,0)$, $E=(t-yt^\nu,t+yt^\nu)$ and consider the joint distribution of
$t^{-1/3}(L(t)-2t)$ and $t^{-1/3}(L(S,E)-2t)$. If $\nu>2/3$, the two random
variables become independent as $t\to\infty$ and if $\nu<2/3$ the joint
distribution is concentrated on the diagonal. Only for $\nu=2/3$ there is a
non-degenerate joint distribution which can be written in terms of suitable
determinants involving the Airy operator $-\frac{d^2}{dx^2}+x$ on $L^2(\R)$.

In our present work we plan to study a related, but more geometrical quantity,
see Figure~\ref{simulation} which displays the directed polymers rotated by
$\pi/4$ for better visibility. The root point is always $S=(0,0)$ and the
end points lie on the line $U_t=\{(t-x,t+x), \vert x \vert \leq t\}$. For fixed
realization $\omega$ and for each end point $E$ we draw the set of all
maximizers. Note that, e.g. for $E=(t,t)$, the directed polymer splits and
merges again, which reflects that $\Pi_{\max}(S,E,\omega)$ contains many paths, their
number presumably growing exponentially in $t$. The resulting network of lines has
some resemblance to a river network with $(0,0)$ as the mouth or to a system of
blood vessels, see~\cite{Me} for related models. To characterize the network a natural
geometrical object is the \emph{last branching} for a pair of directed polymers
with distinct end points~\cite{FH}. As in Figure~\ref{simulation} the starting point is
always $S=(0,0)$ and the end point $E$ must lie on the line $U_t$. If $\pi_i$ is
a maximizer with start point $S=(0,0)$ and end point $E_i \in U_t$, $i=1,2$,
then the last point in which $\pi_1$ and $\pi_2$ intersect is denoted by
$I(\pi_1,\pi_2)$. We define the \emph{last intersection point} for two sets of maximizers
by $$J(E_1,E_2)= I(\pi_1,\pi_2)\textrm{ which minimizes }d(I(\pi_1,\pi_2),U_t)$$
where $d(X,U_t)$ is the Euclidean distance between $X$ and $U_t$.
$J(E_1,E_2)$ depends on the configuration $\omega$ of the Poisson points but is
independent from the choice of the maximizers. $J(E_1,E_2)$ is unique since
the existence of two distinct last intersection points is in
contradiction with the condition of being the last intersection. In particular
if $(E_1)_1< (E_2)_1$, then $J(E_1,E_2)$ can be obtained by taking the highest
maximizer from $0$ to $E_2$ and the lowest maximizer from $0$ to $E_1$.

Instead of the geometrical intersection, one could require the last
intersection point to be a Poisson point. The two maximisers have then
necessarily a common root. For the coarse quantities studied here there
is no distinction and our results are identical in both cases.

One would expect that the branching is governed again by the transverse exponent
$2/3$. More precisely let us assume that $$d(E_1,E_2)=\Or(t^\nu), \,0\leq \nu <1
\textrm{ and }E_1,E_2 \in U_t.$$
If $\nu=2/3$, the last branching point should have a distance of order $t$ from
$U_t$ with some on that scale non-degenerate distribution. On the other hand if
$\nu>2/3$ the branching will be close to the root and if $\nu<2/3$ the branching
will be close to $U_t$. Our main result is to indeed single out $\nu=2/3$ and
provide some estimates on the tails.

\begin{thm}\label{thm1}
Let $E_1=(t,t)$ and $E_2=E_1+y t^\nu (-1,1)$ with $y \sim \Or(1)$.
\begin{itemize}
\item[i)] For $\nu>2/3$, there exists a $C(y)<\infty$ such that for all
$\sigma>5/3-\nu$,
$$\lim_{t\to\infty}\Pb(\{d(0,J(E_1,E_2))\leq C(y) t^\sigma\})=1.$$
\item[ii)] For $\nu\leq 2/3$ and for all $\mu<2\nu-1/3$ one has
$$\lim_{t \to \infty}\Pb(\{d(J(E_1,E_2),U_t)\leq t^\mu\})=0.$$
\end{itemize}
In particular for $\nu=2/3$, one can choose $\mu<1$.
\end{thm}

Our result does not rule out the possibility that for $\nu=2/3$ the
distribution of the last intersection point is degenerate near the origin. In
fact the proof exploits geometric aspects for branching points close to $U_t$,
which cannot be used to obtain sharp results close to the origin.

Another way to characterize the network of Figure~\ref{simulation} is to
consider the line density at the cross-section $U_s$, equivalently the typical
distance between maximizers when crossing $U_s$. To have a definition, for given
$\omega$ let $\overline{M}_t$ be all the maximizers with end points in $U_t$
considered as a subset of $\{(x_1,x_2), 0\leq x_1+x_2\leq 2t\}$.
$\overline{M}_t$ consists of straight segments connecting two points of $\omega$
and straight segments connecting $0$ with a point in $\omega$. In addition there
is a union of triangles with base contained in $U_t$ and the apex a point of
$\omega$. We define $M_t$ to be $\overline{M}_t$ such that in every triangle
only the two sides emerging from the apex are retained. Let
\begin{equation}\label{Nt}N_t(s)=\#\textrm{ of points of }M_t\cap U_s,\, 0 \leq s \leq t.\end{equation}
If $s=c\, t$, $0<c<1$, then the typical distance between lines is of order
$t^{2/3}$ and thus one expects $$N_t(ct)\simeq t^{1/3}.$$
On the other hand for a cross-section closer to $U_t$ the number of points
should increase faster. In particular $N_t(t)\simeq t$. This suggests that
$$N_t(t-t^\mu)\simeq t^{g(\mu)}$$
with $g(0)=1$ and $g(1)=1/3$. In the last section we prove the lower bound
$$g(\mu)\geq \frac{5}{6}-\frac{\mu}{2}.$$

\section{Last branching}
We plan to prove Theorem~\ref{thm1}. Before we introduce some notation and
state some results of \cite{BDJ} concerning large deviations for the length of maximizers.

For any $w\prec w' \in \R_+^2$, we denote by $[w,w']$ the rectangle with corners
at $w$ and $w'$ and by $a(w,w')$ its area.
The maximal length $L(w,w')$ is a random variable whose distribution
function depends only on $a(w,w')$ with $L(w,w')\sim 2\sqrt{a(w,w')}$.
Large deviation estimates for $\Pb(L(w,w')\leq 2\sqrt{a(w,w')} + n)$
are proved in \cite{BDJ}, Lemma 7.1. We consider the case of $a(w,w')\gg 1$ and
$\vert n \vert \gg 1$. Let $$\tau=n (\sqrt{a(w,w')}+n/2)^{-1/3}.$$ 
Then there are some positive constants $\theta, T_0, c_1, c_2$ so that\\[12pt]
1. {\it Upper tail}: if $T_0\leq \tau$ and $n\leq 2\sqrt{a(w,w')}$, then
\begin{equation}
\Pb(L(w,w')\geq 2\sqrt{a(w,w')}+n)\leq c_1 \exp(-c_2 \tau^{3/2}),
\label{upLD}\end{equation}\\
2. {\it Lower tail}: if $\tau\leq -T_0$ and $\vert n \vert \leq 2\sqrt{a(w,w')} \,\theta$, then 
\begin{equation}
\Pb(L(w,w')\leq 2\sqrt{a(w,w')}-\vert n \vert) \leq c_1 \exp(-c_2\vert{\tau}\vert^3).
\label{lowLD}\end{equation}

Our first step is to prove a lemma on the length (as in [Jo], Lemma 3.1) and a geometric
lemma because both will be used in the proofs.
Let $\mathcal{Z}$ be a set of points in $\R_+^2$ such that
$\vert\mathcal{Z}\vert \leq t^m$ for a finite $m$ and let for each
$z\in\mathcal{Z}$ be $z'=z+\hat{x}$ where $\hat{x}$ is a unit vector of
$\R_+^2$.
\begin{lem}\label{lemmalength} Let $\delta\in(1/3,1)$ and $E$ a fixed end point
on $U_t$. For each $z\in \mathcal{Z}$, $$E_z=\{\omega \in \Omega \st
L(0,z')\leq 2 \sqrt{a(0,z')}+t^\delta \textrm{ and }L(z,E)\leq 2
\sqrt{a(z,E)}+t^\delta\}.$$ Then for all $\e>0$ and $t$ large
enough we have $$\Pb\bigg(\bigcup_{z\in\mathcal{Z}}\Omega\setm
E_z\bigg)\leq \e.$$
\begin{proof}[\sc{Proof}]
Let $\omega([z,E])$ be the number of Poisson points in $[z,E]$.
If $a=a(z,E)\leq t^{\delta/2}$, then $$\Pb(\omega([z,E])\geq t^{\delta}) 
=\sum_{j\geq t^{\delta}}{\eone^{-a}\frac{a^j}{j!}}\leq C \sum_{j\geq
t^\delta}{\eone^{-a f(j/a)}},$$ where $f(x)=1-x+x\ln{x}$ and $C>0$ a constant
(using Stirling's formula). But for $x>7$, $f(x)>x$ and here $x=j/a\geq t^{\delta/2}\gg 1$,
therefore \begin{equation}\label{lem2eq1}\Pb(L(z,E)\geq
2\sqrt{a(z,E)}+t^\delta)\leq \Pb(\omega([z,E])\geq t^{\delta})\leq C \sum_{j\geq
t^\delta}{\eone^{-j}} \leq 2 C \eone^{-t^{\delta}}.\end{equation} The same bound holds for
$\Pb(L(0,z')\geq 2 \sqrt{a(0,z')}+t^\delta)$.

If $a=a(z,E)\geq t^{\delta/2}$ then
$$\Pb(L(z,E)\geq 2\sqrt{a}+t^\delta) \leq
\Pb(L(z,E)\geq 2\sqrt{a}+a^{\delta/2}).$$
Consequently taking $n=a^{\delta/2}$, we have
$\tau=a^{\delta/2-1/6}(1+o(1))$. Moreover $\tau\geq t^{(\delta-1/3)\delta/4}/2$
for $t$ large enough, because $a\geq t^{\delta/2}$ and consequently by (\ref{upLD})
\begin{equation}\label{lem2eq2}\Pb(L(z,E)\geq 2\sqrt{a(z,E)}+t^\delta) \leq
c_1 \exp(-c_2 t^{(3\delta-1)\delta/8}/3).\end{equation}
The same estimate holds for $\Pb(L(0,z')\geq 2\sqrt{a(0,z')}+t^\delta)$.
Since $-t^\delta\ll -t^{(3\delta-1)\delta/8}$ for $t$ large,
combining (\ref{lem2eq1}) and (\ref{lem2eq2}) we have
\begin{equation}\label{EndLemma2}
\Pb\bigg(\bigcup_{z\in\mathcal{Z}}\Omega\setm E_z\bigg)
\leq t^m \max_{z\in\mathcal{Z}}\Pb(\Omega\setm E_z)\leq t^m c_1 \exp(-c_2
t^{(3\delta-1)\delta/8}/3)\leq \e
\end{equation} for $t$ large enough.
\end{proof}
\end{lem}

\begin{figure}[t]
\begin{center}
\psfrag{A}[][][1.8]{$A$}
\psfrag{B}[][][1.8]{$B$}
\psfrag{A'}[][][1.8]{$A'$}
\psfrag{B'}[][][1.8]{$B'$}
\psfrag{u}[][][1.8]{$U_t$}
\psfrag{w}[][][1.8]{$w$}
\psfrag{l}[][][1.8]{$l$}
\psfrag{O}[][][1.8]{$0$}
\psfrag{E}[][][1.8]{$E$}
\psfrag{z_j}[][][1.8]{$z_j$}
\psfrag{Cyl}[][][1.8]{$C(w,l)$}
\psfrag{1}[][][1.8]{$1$}
\psfrag{2}[][][1.8]{$2$}
\includegraphics[angle=0,width=8cm]{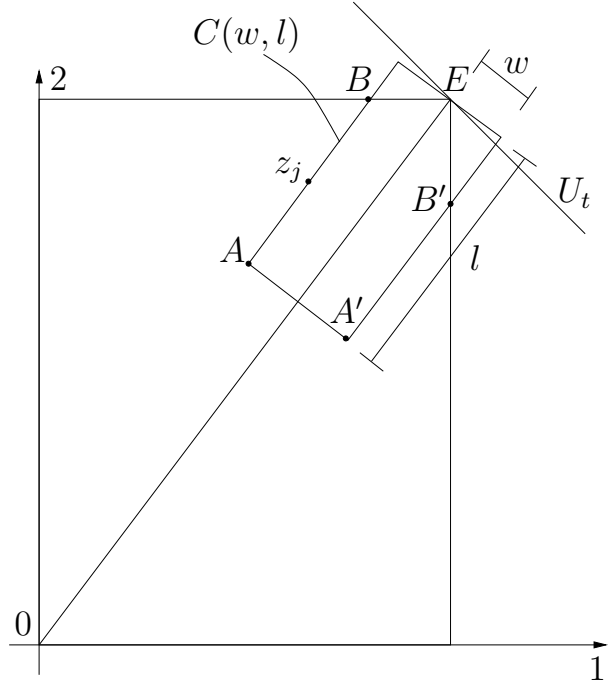}
\caption{\it Geometrical construction used in Lemma~\ref{lemmageom} and
Theorem~\ref{thm1} $i)$.}\label{fig1}
\end{center}
\end{figure}

Let us consider an end point $E$ on $U_t$ given by $E=(t(1-k),t(1+k))$ with $k\in
(-1,1)$ and let $\hat{x}$ be the unit vector with direction $\overrightarrow{0E}$.
The cylinder $C(w,l)$ has axis $\overline{0E}$, width $w$ and length $l$ (see
Figure~\ref{fig1}). $\partial C(w,l)$ is the boundary of the cylinder without lids. Then
the following geometric lemma holds.

\begin{lem}\label{lemmageom}
Let $z\in\partial C(w,l)$ with $w=t^\nu$, $l=t^\mu$, $\nu<\mu$ and
$z'=z+\hat{x}$. Then there exists a $C(k)>0$ such that
\begin{equation}\label{eqlemgeom}
\sqrt{a(0,z')}+\sqrt{a(z,E)}-\sqrt{a(0,E)}\leq -C(k) \frac{w^2}{l}\textrm{ as
}t\to\infty.
\end{equation}
\begin{proof}[\sc{Proof}]
First let us consider $\mu<1$.
Let $e_1= \frac{1}{\sqrt{2(1+k^2)}}\binom{1-k}{1+k}$ and
$e_2=\frac{1}{\sqrt{2(1+k^2)}}\binom{-(1+k)}{1-k}$. Then $z=E\pm w e_2 -
\lambda l e_1$ with $\lambda\leq 1$ such that $z\in [0,E]$. For the
computations we consider the "$+$" case, the "$-$" case is obtained replacing
$w$ with $-w$ at the end. Let $Q= \sqrt{2(1+k^2)}$ and $l'= \lambda l
-1$. Then
$$z=\binom{t(1-k)-w (1+k)/Q-\lambda l (1-k)/Q}{t(1+k)+w(1-k)/Q-\lambda l (1+k)/Q}$$
and $$z'=\binom{t(1-k)-w (1+k)/Q-l' (1-k)/Q}{t(1+k)+w(1-k)/Q-l'(1+k)/Q}.$$
Expansion leads to the following results,
\begin{eqnarray*}\sqrt{a(0,z')}&=&t \sqrt{1-k^2}-\frac{\lambda
l}{\sqrt{2}}\sqrt{\frac{1-k^2}{1+k^2}}-\frac{\sqrt{2}k
w}{\sqrt{1-k^4}}+\Or(w^2/t),\\
\sqrt{a(0,E)}&=&t\sqrt{1-k^2},\\
\sqrt{a(z,E)}&=&\frac{\lambda l}{\sqrt{2}}\sqrt{\frac{1-k^2}{1+k^2}}
f(k,w/\lambda l),
\end{eqnarray*} where
$f(k,\zeta)=\sqrt{1+4 k \zeta (1-k^2)^{-1}-\zeta^2}$.
It follows that
$$\sqrt{a(0,z')}+\sqrt{a(z,E)}-\sqrt{a(0,E)}=-\sqrt{\frac{1-k^2}{2(1+k^2)}}
h(k,w,\lambda l)+\Or(w^2/t),$$ where
$$h(k,w,\lambda l)=\lambda l \bigg(1-\sqrt{1+\frac{4 k w}{\lambda l
(1-k^2)}-\frac{w^2}{\lambda^2 l^2}}+\frac{2 k w}{\lambda l (1-k^2)}\bigg).$$
It is easy to see that $h(k,w,\lambda l)\geq 0$ (in fact, $h(k,w,\lambda
l)=0$ only if $\frac{1}{\lambda l}=0$). Moreover $$h(k,w,\lambda l)\sim
\frac{(k^2+1)^2 w^2}{2(k^2-1)^2 \lambda l} + \Or(w^3/(\lambda l)^2),$$ and
the minimal value is obtained for $\lambda=1$. Consequently, for $l=t^\nu$ large
enough,
$$\sqrt{a(0,z')}+\sqrt{a(z,E)}-\sqrt{a(0,E)}\leq - C(k)\frac{w^2}{l}\textrm{
with }C(k)=\frac{(k^2+1)^2}{4(k^2-1)^2}.$$

Secondly let us consider the case $\mu=1$. In this case a $z\in\partial C(w,l)$ can be written as
$$z=\binom{\alpha t(1-k)-w (1+k)/Q}{\alpha t(1+k)+w(1-k)/Q}\textrm{ and }
z'=\binom{\alpha' t(1-k)-w (1+k)/Q}{\alpha' t(1+k)+w(1-k)/Q}$$ where
$\alpha \in (0,1)$ such that $z\in [0,E]$ and $\alpha'=\alpha+1/tQ$. The
expansion yields to
\begin{eqnarray*}
\sqrt{a(0,z')}+\sqrt{a(z,E)}-\sqrt{a(0,E)}&=&
-\left(\frac{1}{\alpha}+\frac{1}{1-\alpha}\right)
\frac{1+k^2}{4(1-k^2)^{3/2}}\frac{w^2}{t}+\Or(w^3/t^2)\\
\leq -\frac{1+k^2}{(1-k^2)^{3/2}}\frac{w^2}{t}+\Or(w^3/t^2) &\leq& -C(k)
\frac{w^2}{t}\textrm{ with }C(k)=\frac{1+k^2}{2(1-k^2)^{3/2}}
\end{eqnarray*}
for $t$ large enough.
\end{proof}
\end{lem}

\noindent {\bf Proof of Theorem~\ref{thm1}:}
\begin{proof}[{\sc Proof of} $i)$]
Let $E_1=(t,t)$ and $E_2=E_1+yt^\nu(-1,1)$ with $\nu>2/3$. First we prove that
for a $E=(t(1-k),t(1+k))$ with $k\in (-1,1)$, all maximizers from $0$ to $E$ are
contained in a cylinder $C(w)$ of axis $\overline{OE}$, width $w=t^{\kappa}$, $\kappa>2/3$, with
probability one (as in Section 3 of~\cite{Jo}). Then we compute the intersection
of such cylinders starting at $0$ and ending at $E_1$ and $E_2$ respectively.

Let us consider the following event:
$$D\equiv D(w)= \{\omega \in \Omega \st \forall\, \pi \in
\Pi_{\max}(0,E,\omega) \textrm{ we have }\pi\cap\partial C(w) = \emptyset\}.$$
We prove that
\begin{equation}\label{eqA1}
\forall \, \e>0, \Pb(D)\geq 1-\e \textrm{ for }t\textrm{ large enough.}
\end{equation}
If $\omega \in \Omega \setm D$, then there exists a maximizer $\pi$ such that
$\pi\cap\partial C(w)\neq \emptyset$. We divide the two sides of 
$\partial C(w)$ in $K=2t$ equidistant points (see Figure~\ref{fig1}) with
$A=z_0$, $B=Z_K$ and $z_j=A+j\vert AB \vert/K \hat{x}$ where $\hat{x}$ is the
unit vector with
direction $\overrightarrow{0E}$. Likewise for the second side of the cylinder. Let
$\mathcal{A}$ be the set of all these points. We define $z(\omega)$ as follows:
if the last intersection of $\pi$ with $\partial C(w)$ is in
$\overline{z_{j-1} z_j}$, then $z(\omega)=z_j$ (with $z(\omega)=z_j$ is the
intersection is exactly at $z_j$), and $z'(\omega)=z(\omega)+\hat{x}$. Then we
have 
\begin{equation}\label{eqA2}
L(0,E)\leq L(0,z'(\omega))+L(z(\omega),E).\end{equation}
Defining for all $z\in\mathcal{A}$
\begin{equation}\label{eqA3}
E_z=\{\omega \in \Omega \st L(0,z')\leq 2 \sqrt{a(0,z')}+t^\delta
\textrm{ and }L(z,E)\leq 2 \sqrt{a(z,E)}+t^\delta\}
\end{equation}
we obtain, by Lemma~\ref{lemmalength}, that for $\delta>1/3$
$$\Pb\bigg(\bigcup_{z\in\mathcal{A}}\Omega\setm E_z\bigg)\leq
\e\textrm{ for all }\e>0\textrm{ and }t\textrm{ large enough.}$$
We consider now the set of events
$F= (\Omega\setm D)\bigcap_{z\in\mathcal{A}}E_z$.
Then $\Pb(F)=1-\Pb(\Omega\setm F)\geq 1-\Pb(D)-
\Pb(\bigcup_{z\in\mathcal{A}}\Omega\setm E_z) \geq \Pb(\Omega\setm D) - \e$
if $t$ is large enough, that means $$\Pb(\Omega\setm D)\leq \e + \Pb(F).$$
We need to prove that $\Pb(F)\leq \e$ for $t$ large enough. For all $\omega \in
F$, from (\ref{eqA2}) and (\ref{eqA3}) follows:
\begin{equation}\label{eqA4}
L(0,E)\leq 2 t^\delta + 2 (\sqrt{a(0,z'(\omega))}+\sqrt{a(z(\omega),E)}).
\end{equation}
Applying Lemma~\ref{lemmageom} with $\mu=1$ we obtain
\begin{equation}\label{eqA5}
\sqrt{a(0,z'(\omega))}+\sqrt{a(z(\omega),E)}\leq \sqrt{a(0,E)}-C(k) t^{2\kappa-1}.
\end{equation}
From (\ref{eqA4}) and (\ref{eqA5}) we have, for all $\omega \in F$,
$L(0,E)-2\sqrt{a(0,E)}\leq 2 t^\delta - 2C(k) t^{2\kappa-1}$ for $t$ large enough.
This implies, taking $\delta<2\kappa-1$ (always possible since $2\kappa-1>1/3$), for $t$ large enough,
\begin{eqnarray*}
\Pb(F)&\leq&\Pb(L(0,E)-2\sqrt{a(0,E)}\leq 2 t^\delta - 2C(k) t^{2\kappa-1})\\
&\leq & \Pb(L(0,E)-2\sqrt{a(0,E)}\leq  - C(k) t^{2\kappa-1}) \leq \e
\end{eqnarray*}
because $- t^{2\kappa-1}/t^{1/3}\to -\infty$ as $t\to\infty$. This proves
(\ref{eqA1}).

Therefore with probability approaching to one as $t$ goes to infinity,
the maximizers from $0$ to $E$ are in a cylinder of width $w=t^\kappa$ with
$\kappa>2/3$. We use the result for $E=E_1$ and for $E=E_2$.
Let us take $\kappa \in (2/3,\nu)$ and let $C_1, C_2$ be the cylinders that include
the maximizers from $0$ to $E_1, E_2$ respectively. Let $G$ be the farthest
point from the origin in $C_1\cup C_2$. Then for $t$ large enough and for all $\e>0$,
\begin{equation}\label{eqA6}
\Pb(d(0,J(\omega))\leq d(0,G))\geq 1-2\e.
\end{equation}
We need only to compute $d(0,G)$. By some algebraic computations we obtain
$$d(0,G)=\frac{t^{\kappa+1-\nu}}{\vert y\vert}+\Or(t^{\nu-1}) \leq
\frac{2t^{\kappa+1-\nu}}{\vert y\vert}\textrm{ for }t\textrm{ large enough}$$
and $\kappa \in (2/3,\nu)$ implies $\kappa+1-\nu>5/3-\nu$.
\end{proof}

\begin{proof}[{\sc Proof of} $ii)$]
We consider the case $y>0$, the case $y<0$ is obtained by symmetry.
Let us consider the cylinder $C(w,l)$ with axis $\overline{0 E_1}$ of length $l=t^\mu$
and width $w=yt^\nu$, $\nu<\mu$. We note by $\partial C(w,l)_{+}$ the upper side of
$C(w,l)$ (see Figure~\ref{fig2}).

\begin{figure}[t]
\begin{center}
\psfrag{A}[][][1.8]{$A$}
\psfrag{B}[][][1.8]{$B$}
\psfrag{u}[][][1.8]{$U_t$}
\psfrag{O}[][][1.8]{$0$}
\psfrag{E_1}[][][1.8]{$E_1$}
\psfrag{E_2}[][][1.8]{$E_2$}
\psfrag{d_m}[][][1.8]{$d_m$}
\psfrag{Cyl}[][][1.8]{$C(w,l)$}
\psfrag{1}[][][1.8]{$1$}
\psfrag{2}[][][1.8]{$2$}
\includegraphics[angle=0,width=8cm]{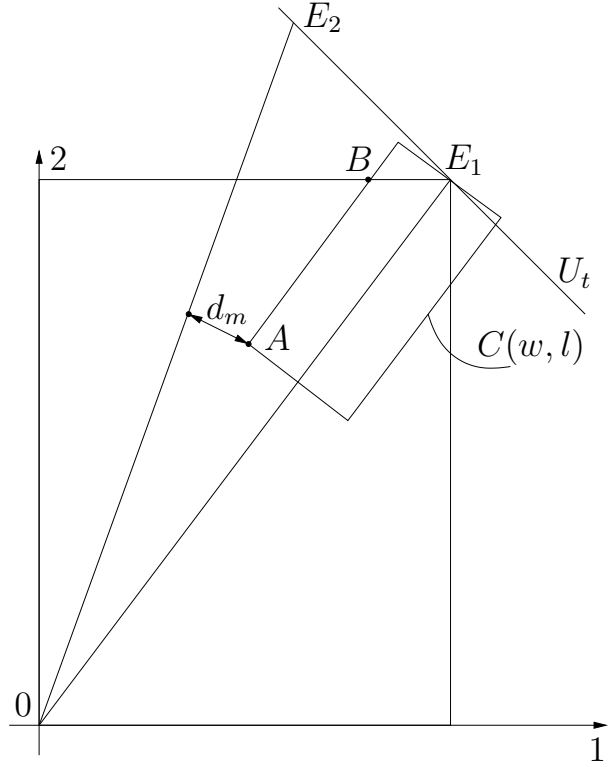}
\caption{\it Geometrical construction used in Theorem~\ref{thm1} $ii)$ and in
Theorem~\ref{thm2}.}\label{fig2}
\end{center}
\end{figure}

Let $$D\equiv D(w,l)= \{\omega \in \Omega \st \forall\, \pi \in
\Pi_{\max}(0,E_1,\omega) \textrm{ we have }\pi\cap\partial C(w,l)_{+} = \emptyset\}.$$
If $\omega \in \Omega \setm D$ then the highest maximizer, $\pi_0$, from $0$ to
$E_1$ intersect $\partial C(w,l)_{+}$ in $\overline{AB}$. We divide $\overline{AB}$ in
$K=[\sqrt{2}(l-w)]+1$ equidistant points with $A=z_0$, $B=z_K$ and
$z_j=A+j(l-w)(1,1)/K$.
Let $\mathcal{A}$ be the set of all these points. We define $z(\omega)$ as
follows: if the last intersection of $\pi_0$ with $\partial C(w,l)_{+}$ is in
$\overline{z_{j-1} z_j}$, then $z(\omega)=z_j$ (with $z(\omega)=z_j$ if the
intersection is exactly at $z_j$) and $z'(\omega)=z(\omega)+(1,1)/\sqrt{2}$.
We have
\begin{equation}\label{eqB1}
L(0,E_1)\leq L(0,z'(\omega))+L(z(\omega),E_1).
\end{equation}
We define for all $z\in\mathcal{A}$
\begin{equation}\label{eqB2}
E_z=\{\omega \in \Omega \st L(0,z')\leq 2 \sqrt{a(0,z')}+t^\delta
\textrm{ and }L(z,E_1)\leq 2 \sqrt{a(z,E_1)}+t^\delta\}
\end{equation}
and the set of events $F= (\Omega \setm D)\bigcap_{z\in\mathcal{A}}E_z$.
In what follows we consider $\delta>1/3$.
Then using Lemma~\ref{lemmalength} we conclude that for all $\e>0$ and $t$
large enough
$$\Pb(\Omega\setm D)\leq \e + \Pb(F).$$
For all $\omega \in F$, from (\ref{eqB1}) and (\ref{eqB2}) follows:
\begin{equation}\label{eqB3}
L(0,E_1)\leq 2 t^\delta + 2(\sqrt{a(0,z'(\omega))}+\sqrt{a(z(\omega),E_1)}).
\end{equation}
From the geometric Lemma~\ref{lemmageom} we deduce
\begin{equation}\label{eqB4}
\sqrt{a(0,z'(\omega))}+\sqrt{a(z(\omega),E_1)}\leq\sqrt{a(0,E_1)}-C y^2 t^{2\nu-\mu}.
\end{equation}
Therefore for all $\omega \in F$, by (\ref{eqB3}) and (\ref{eqB4}),
$L(0,E_1)-2t\leq 2 t^\delta - 2C y^2 t^{2\nu-\mu}\leq -C y^2 t^{2\nu-\mu}$ for
$t$ large enough if $2\nu-\mu>\delta$. This implies that for all $\e>0$ and
$t$ large enough
$$\Pb(\Omega \setm D)\leq \e+\Pb(F)\leq \e+ \Pb(L(0,E_1)-2t\leq -C y^2
t^{2\nu-\mu})\leq 2\e \textrm{, if }\mu<2\nu-\delta.$$
Let now define the set of events $$Q= \{\omega \in \Omega \st
d(J(E_1,E_2)(\omega),u)\leq l\textrm{ with }l=t^{\mu}\}.$$ We need to prove
that 
\begin{equation}\label{eqB5}
\lim_{t\to\infty}\Pb(Q)=0\textrm{ for all }\mu<2\nu-1/3.
\end{equation}
We consider the event $T= Q \cap D$ with $\mu<2\nu-1/3$. For any choice
of $\mu<2\nu-1/3$, there exists a $\delta>1/3$ such that $\mu<2\nu-\delta$ is
verified. Then for $t$ large enough we have
$\Pb(T)\geq \Pb(D)+\Pb(Q)-1\geq \Pb(Q)-\e$, i.e. $\Pb(Q)\leq \Pb(T)+\e$.\\
If $\omega \in T$ then the lowest maximizer from $0$ to $E_2$ intersect
$\partial C(w,l)_{+}$ at some point $H$,
$$H=\binom{t}{t}-\frac{\lambda}{\sqrt{2}}\binom{l}{l}+\frac{1}{\sqrt{2}}\binom{-w}{w}$$
with $\lambda \in (0,1]$ such that $(H)_1\leq t-yt^{\nu}$. We define
$h(\omega)=z_j\in\mathcal{A}$ if $H(\omega)\in \overline{z_{j-1}z_j}$ (always
with $h(\omega)=z_j$ if $H(\omega)=z_j$) and $h'(\omega)=h(\omega)+(1,1)/\sqrt{2}$. As before, for $\omega \in T$ we have
$$L(0,E_2)\leq 2 t^\delta+2(\sqrt{a(0,h'(\omega))}+\sqrt{a(h(\omega),E_2)}).$$
In order to apply the geometric lemma we need to know the minimal distance
$d_m$ between $\partial C(w,l)_{+}$ and the segment $\overline{0 E_2}$.
We find $d_m=(\sqrt{2}-1)yt^\nu+\Or(t^{\nu+\mu-1})$.\\
Applying Lemma~\ref{lemmageom} we obtain
$$\sqrt{a(0,h'(\omega))}+\sqrt{a(h(\omega),E_2)}\leq\sqrt{a(0,E_2)}-C'
t^{2\nu-\mu}$$ provided that $\mu<2\nu-1/3$.
Therefore for all $\e>0$ and $\mu<2\nu-1/3$, $$\Pb(Q)\leq\Pb(T)+\Pb(\Omega\setm
D)\leq \Pb(L(0,E_2)-2\sqrt{a(0,E_2)}\leq -C' t^{2\nu-\mu})+
2\e\leq 3\e$$ for $t$ large enough.
\end{proof}

\section{Density of branches}
We recall the definition (\ref{Nt}) of the number of branches $N_t(s)$ at
cross-section $U_s$.

\begin{thm}\label{thm2}
For $0\leq\mu<1$ the following lower bound holds,
$$\lim_{t\to\infty}\Pb\left(N_t(t-t^\mu)\geq t^\sigma\right)=1$$
for all $\sigma<\frac{5}{6}-\frac{\mu}{2}.$
\begin{proof}[{\sc Proof}]
The first part of the proof is close to the one of Theorem~\ref{thm1} ii).

As in Figure~\ref{fig2} let us consider two fixed points
$$E_1=(t(1-k),t(1+k))\textrm{ and }E_2=E_1+t^\nu (-1,1)$$ with $k\in(-1,1)$. We
look at the region closer than $l=t^\mu$ from the line $U_t$.  We take
$w=t^\nu/2$ and define $C(w,l)$, $\partial C(w,l)_+$ and $D=D(w,l)$ as in
the previous proof. We divide $\overline{AB}$ in $K=[\vert AB\vert]+1$
equidistant points $z_j$,
define the $z(\omega)$, $z'(\omega)$ and $E_z$ (see (\ref{eqB2})) as in the previous proof.
Equation (\ref{eqB1}) holds unchanged too.
Let $F= (\Omega \setm D)\bigcap_{z\in\mathcal{A}}E_z$.
Then for $\delta>1/3$ the proof of Lemma~\ref{lemmalength} gives
also (see (\ref{EndLemma2})) $$\Pb\bigg(\bigcup_{z\in\mathcal{A}}\Omega\setm E_z\bigg)\leq
1/t^2\textrm{ for }t\textrm{ large enough.}$$ Then $$\Pb(\Omega\setm D)\leq
t^{-2} + \Pb(F)$$ for $t$ large enough.

We still have (\ref{eqB3}) for all $\omega\in F$ and (\ref{eqB4}) becomes
$$\sqrt{a(0,z'(\omega))}+\sqrt{a(z(\omega),E_1)}\leq \sqrt{a(0,E_1)}-C(k)t^{2\nu-\mu}/4.$$
Therefore, taking $2\nu-\mu>\delta$, for $t$ large enough
$$\Pb(F)\leq \Pb(L(0,E_1)\leq 2\sqrt{a(0,E_1)}-C(k) t^{2\nu-\mu}/4).$$
Let $\psi= \min\{2\nu-\mu,(1+\delta)/2\}$, then 
$$\Pb(F)\leq \Pb(L(0,E_1)\leq 2\sqrt{a(0,E_1)}-C(k)t^\psi/4).$$
The $\psi$ is introduced in order to remain in the domain in which (\ref{lowLD})
can be applied. The large deviation estimate leads to
$$\Pb(F)\leq c_1 \exp(-c_2 P(k) t^{3\psi-1})\textrm{ with }
P(k)=\frac{C(k)^3}{4^3 2\sqrt{1-k^2}}>0.$$
Therefore
$\Pb(F)\leq 1/t^2$ for $t$ large enough. Consequently
$$\Pb(\Omega\setm D)\leq t^{-2}+\Pb(F)\leq 2t^{-2}.$$

Define the set of events $$Q= \{\omega \in \Omega \st
d(J(E_1,E_2)(\omega),U_t)\leq l\textrm{ with }l=t^{\mu}\}.$$ We prove that for
$t$ large enough
\begin{equation}
\Pb(Q)\leq t^{-2}\textrm{ for all }\mu<2\nu-1/3.
\end{equation}
We consider the event $T= Q \cap D$ with $\mu<2\nu-1/3$. As in the previous
proof, for $t$ large enough we have $\Pb(Q)\leq \Pb(T)+\Pb(\Omega\setm D)$.
If $\omega \in T$ then the lowest maximizer from $0$ to $E_2$ intersects
$\partial C(w,l)_{+}$ at some point $H$. We define $h(\omega)$ and $h'(\omega)$
as in the previous proof and for $\omega \in T$ we have
$$L(0,E_2)\leq 2 t^\delta+2(\sqrt{a(0,h'(\omega))}+\sqrt{a(h(\omega),E_2)}).$$
We compute the minimal distance $d_m$ between $\partial C(w,l)_{+}$ and the segment $\overline{0 E_2}$
finding
$d_m=\left(\frac{\sqrt{2}}{\sqrt{1+k^2}}-\frac{1}{2}\right)t^\nu+\Or(t^{\mu-1})$.
Applying Lemma~\ref{lemmageom} we obtain
$$\sqrt{a(0,h'(\omega))}+\sqrt{a(h(\omega),E_2)}\leq\sqrt{a(0,E_2)}-C'(k)
t^{2\nu-\mu}/2$$ provided that $\mu<2\nu-1/3$ and with
$C'(k)=C(k)\left(\frac{\sqrt{2}}{\sqrt{1+k^2}}-\frac{1}{2}\right)^2$.
Therefore for \mbox{$\mu<2\nu-1/3$},
$$\Pb(T)\leq \Pb(L(0,E_2)-2\sqrt{a(0,E_2)}\leq -C'(k) t^{2\nu-\mu})
\leq \Pb(L(0,E_2)-2\sqrt{a(0,E_2)}\leq -C'(k) t^{\psi}).$$
Applying (\ref{lowLD}) with $n=-C'(k) t^{\psi}$ we obtain
$$\Pb(T)\leq c_1 \exp(-c_2 C''(k) t^{3\psi-1})$$ with $C''(k)=C'(k)^3/2\sqrt{1-k^2}$.
Therefore $\Pb(T)\leq 1/t^2$ for $t$ large enough. Finally for $t$ large enough
$$\Pb(T)\leq \Pb(\Omega\setm D)+\Pb(T)\leq 3/t^2$$
provided that $\mu<2\nu-1/3$.

Now we can prove the theorem. Let us fix $0<k_0\ll 1$ and
$M=[(1-k_0)t^{1-\nu}]$. We choose $2M+1$ points on $U_t$ as follows: $T_0=(t,t)$
and $T_j=T_0+j t^\nu (-1,1)$ for $j=-M,\ldots,M$. Let $W(j)$ be the set of all
intersections between the maximizers with end point at $T_j$ and the ones with
end point at $T_{j+1}$. We define $m(j)$ to be the set of points of $W(j)$ whose distance
to $U_t$ is at most $l=t^\mu$. Then
$$\Pb(\exists \, j \st m(j)\neq \emptyset)=\Pb\bigg(\bigcup_{j=-M}^{M-1}m(j)\neq
\emptyset\bigg)\leq 2 M \max_{j=-M,\ldots,M-1}\Pb(m(j)\neq \emptyset) \leq 6
t^{-1-\nu}$$ as $t$ goes to infinity. Then as $t$ goes
to infinity we have, at distance $t^\mu$ with $\mu<2\nu-1/3$, 
at least $2M+1\sim t^{1-\nu}$ branches that have not yet merged with probability
one. Since for $k_0\ll 1$, $2M+1 =2[(1-k_0)t^{1-\nu}]+1\geq t^{1-\nu}$, 
for all $\sigma= 1-\nu<5/6-\mu/2$ we have
$$\lim_{t\to\infty}\Pb\left(N_t(t-t^\mu)\geq t^\sigma\right)=1.$$
\end{proof}
\end{thm}

\section*{Acknowledgments}
We thank Michael Pr\"ahofer for helping us to generate Figure~\ref{simulation}.

\end{document}